\documentclass[11pt]{article}

\pdfoutput=1

\usepackage[utf8]{inputenc}

\usepackage{xcolor} %
\usepackage{soul} %

\usepackage[style=numeric,
backend=biber,
doi=false,
isbn=false,
date=year,
giveninits=true]{biblatex}

\DeclareNameAlias{default}{family-given}

\DefineBibliographyStrings{english}{%
 andothers = {\addcomma\addspace\textsc{et\addabbrvspace al}\adddot},
 and = {\textsc{and}}
}

\DeclareFieldFormat
  [article,inbook,incollection,inproceedings,patent,thesis,unpublished]
  {title}{#1}

\renewbibmacro{in:}{%
  \ifentrytype{article}{%
  }{%
    \printtext{\bibstring{in}\intitlepunct}%
  }%
}

\renewbibmacro*{volume+number+eid}{%
  \printfield{volume}%
  \setunit*{\addcomma\space}%
  \printfield{number}%
  \setunit{\addcomma\space}%
  \printfield{eid}}

\DeclareFieldFormat{pages}{#1}

\renewbibmacro*{publisher+location+date}{%
  \printlist{publisher}%
  \setunit*{\addcomma\space}%
  \printlist{location}%
  \setunit*{\addcomma\space}%
  \usebibmacro{date}%
  \newunit}

\renewbibmacro*{url+urldate}{%
  \ifboolexpr{
    test {\ifentrytype{inreference}} 
    or 
    test {\ifentrytype{online}}
  }
  {\printfield{url}%
    \iffieldundef{urlyear}
      {}
      {\setunit*{\addspace}%
       \printurldate}}
  {}}

\AtEveryBibitem{%
  \clearfield{note}%
}

\usepackage{hyperref}
\hypersetup{
    colorlinks=true,
    linkcolor=blue,
    filecolor=magenta,      
    urlcolor=cyan,
}

\begin{document}
\def\uline{}

\title{\vskip -1in
Delphic Costs and Benefits in Web Search:
A utilitarian and historical analysis\\
{\large (Preliminary extended abstract)}}

\author{Andrei Z.~Broder \&\ Preston McAfee \\
Google Research\\
Mountain View, CA\\[0.1cm]
{\tt broder@acm.org, prestonmcafee@gmail.com}}

\date{\today}

\maketitle

\begin{abstract}

We present a new framework to conceptualize and operationalize the total user
experience of search, by studying the entirety of a search journey from an
utilitarian point of view.

Web search engines are widely perceived as ``free''. But search requires time and effort:
in reality there are many intermingled non-monetary costs (e.g.~time costs,
cognitive costs, interactivity costs) and the benefits may be marred by various
impairments, such as misunderstanding and misinformation. This characterization
of costs and benefits appears to be inherent to the human search for
information within the pursuit of some larger task: most of the costs and impairments can be identified in interactions
with any web search engine, interactions with public libraries, and even in
interactions with ancient oracles. To emphasize this innate connection, we call
these costs and benefits \textit{Delphic}, in contrast to explicitly financial
costs and benefits.

Our main thesis is that users' satisfaction with a search engine mostly depends
on their experience of Delphic cost and benefits, in other words on their
utility. The consumer utility is correlated with classic measures of search
engine quality, such as ranking, precision, recall, etc., but is not completely
determined by them. To argue our thesis, we catalog the Delphic costs and
benefits and show how the development of search engines over the last quarter
century, from classic Information Retrieval roots to the integration of Large
Language Models, was driven to a great extent by the quest of decreasing Delphic
costs and increasing Delphic benefits.

We hope that the Delphic costs framework will engender new ideas and new
research for evaluating and improving the web experience for everyone.
\end{abstract}

\section*{Introduction}

Twenty five centuries ago, seekers of information rich enough to take a trip to
Mount Parnassus could ask Pythia, the Delphi temple high priestess, whatever
they wanted to know about the past, the present, and the future. Pythias have
been out of business for a long time but, as Julia Kindt has observed in a
\href{https://theconversation.com/friday-essay-secrets-of-the-delphic-oracle-and-how-it-speaks-to-us-today-61738}{\uline{delightful essay}}~\cite{kindt_friday_2016} that inspired this paper, Google and other search engines are modern
substitutes. Kindt noted that \textit{``As with the ancient oracles, answers
provided }[by search engines] \textit{are only ever as good as the question
asked''}.

Here we investigate other similarities: Like the Delphic oracle, web search
engines do not demand any direct payments (and thus are widely perceived as
``free"), but using them requires time and effort. More generally, current
seekers incur costs and obtain benefits in the pursuit of a background task, same as their antique predecessors. Of
course, the modern costs are much lower, but we can identify costs of a similar
type across eons. Furthermore, now as then, different people that make identical
queries and get identical results might experience vastly different costs and
benefits depending on their specific situation: for example, people with good
domain knowledge might value a result that is incomprehensible to those without
it.

Consumers incur many intermingled costs to search: \textit{access costs} (for a
suitable device and internet bandwidth), \textit{cognitive costs} (to formulate
and reformulate the query, parse the search result page, choose relevant
results, etc), \textit{interactivity costs} (to type, scroll, view, click,
listen, and so on), and obviously for all of these activities plus waiting for
results and processing them, \textit{time costs }to \textit{task completion}.

In addition, the  \textit{benefits} of search are highly dependent of the
searcher’s context (e.g.~location, expertise level, prior domain knowledge, {and} their end task) and
might be marred by \textit{impairments}, another type of cost, in the form of
\textit{miscommunication} (the search agent misunderstanding the user intent due, for instance, to user error or inherent ambiguity), 
\textit{misrepresentation} (sites not delivering their promises, e.g.~false
advertising or click bait), \textit{misinformation} (incorrect, outdated, or
incomplete information), \textit{disinformation }(intentional
misinformation) and \textit{misinterpretation} (due e.g.~to user’s
language or reading-level mismatch). Results might have \textit{quality}
\textit{flaws }(e.g.~sites that themselves have large cognitive or interactive
costs, sites that provide sub-par services, etc) and in particular
\textit{safety flaws} (e.g.~sites that misuse or leak private information or
harbor viruses). Other potential impairments to the user experience and/or
reductions in value are \textit{distrust }(the fear, whether founded or not,
that the content presented is flawed, or that the site is unsafe) and
\textit{discomfort} (in reaction to upsetting content).

In addition to access, cognitive, interactivity, and time costs, search presents
a privacy trade-off similar to the choice between using a credit card or paying
cash: cash is more private but cards are more convenient. Searchers inevitably
reveal what they are seeking but can use mechanisms such as history deletion,
anonymous surfing, location hiding, VPN, etc. to reduce the amount of data they
share. These choices however increase other costs and may reduce functionality.
For instance: a query like ``coffee shops near me" is not solvable if the
location is hidden (or increases time and interactivity costs if the location
needs to be made explicit every time); the use of search history improves the
search results; most legitimate VPNs are not free and also increase both latency and access costs.

Turning now to benefits:  Consumers search because they want to complete tasks,  and the queries they make represent steps towards their goals.  Thus the ultimate value provided by search engines is whatever help they provide towards   \textit{task completion}.  However, most of the time the task is unknown and/or the search engines can not offer any  help beyond trying to infer and respond to the query \textit{intent}.   

Classic information retrieval is predicated on queries made due to an \textit{informational need}~\cite{shneiderman_clarifying_1997} necessary for the task at hand, but the intent of web queries  is not always informational: a still popular
\href{https://dl.acm.org/doi/pdf/10.1145/792550.792552}{\uline{2002 taxonomy of
web queries by intent}}~\cite{broder_taxonomy_2002} (devised by one of the authors) identifies 
 three broad categories: (1) \textit{Navigational }(the intent is to reach a particular
site); (2) \textit{Informational} (the intent is to acquire some information);
and (3) \textit{Transactional} (the intent is to perform some web-mediated
activity). For the purposes of this paper it is fitting to carve out (4)
\textit{Hedonic} intent (seeking entertainment, social interaction, or other
forms of leisure directly provided on the web) from other transactional queries,
since in this case the total time to task completion is not necessarily a cost, and the search engine can in fact fulfill the task entirely, rather than provide information that helps a user do it.

Besides online entertainment, there are numerous situations where search engines can go beyond the immediate query intent  to facilitate the task completion.   For instance, a search on Google for a landmark typically offers links to directions; clicking on that link, leads to a map, where  the searcher can book a car service with the pick-up and drop-off addresses already filled.   Compared to cut and paste addresses, and dealing with multiple apps, this scenario offers a significant reduction in the interactivity and time costs required for the corresponding task completion.   Similar examples of such \textit{facilitations} {include} having food or goods delivered, buying tickets to a movie, and making restaurant reservations.

This characterization of total costs and benefits appears to be inherent to the
human search for information: most of the costs can be identified in
interactions with any web search engine, interactions with public libraries, and
even in interactions with ancient oracles. To emphasize this innate connection
we call these costs and benefits \textit{Delphic}, in contrast to explicitly
monetary costs and benefits. Delphic costs are akin to
\href{https://en.wikipedia.org/wiki/Transaction_cost}{\uline{transaction costs}}~\cite{noauthor_transaction_2023}
in economics, but for transactions involving the pursuit of search results
rather than the exchange of goods and services. Our main thesis is that users'
satisfaction with a search engine mostly depends on their experience of Delphic
cost and benefits on their way to task completion, in other words on their utility. This is an instance of a
general economics paradigm: in a first approximation, consumer behavior is well
modeled by assuming consumers are utility maximizers, even though consumers
rarely make a conscious effort to increase their utility. As an example, the
Generation Z penchant to use TikTok as a general search engine might be
explained by the fact that their cognitive processing costs for short videos
(optimized for mobile viewing and their generation’s zeitgeist) is much lower
than their cost of crafting text queries and processing conventional web textual
results.

This ``utilitarian" perspective applies not only to text focused search engines
(e.g.~Bing, Google) but also to video search (e.g.~YouTube, TikTok), product
search (e.g.~Amazon), people search, virtual assistants, etc. 

Most search engines on the web are
supported by advertising. As observed by Phillip Nelson in 1974,
\href{https://www.jstor.org/stable/1837143}{\uline{advertising is a form of
information}}\cite{nelson_advertising_1974}, and hence could be  a consumer benefit insofar as the ads are
relevant to their needs and facilitate their task completion: for instance on a product search, ads specific for that
product typically come from stores that have the product in stock and ads for
related products are based on deeper understanding of the product than what’s
readily ``known" to a general search engine.  Similarly ads often provide deeper (that is, more specific) linking into a
provider site than what the search engine provides thus reducing the interactivity cost. Last, but most
importantly as a  consumer benefit, the existence of ads is what makes search
engines ``free".  Furthermore, the ads themselves are ultimately paid from the money that
people with sufficient disposable income spend on the products being advertised,
thus these people subsidize search for everyone else. On the other hand,
irrelevant ads are a distraction and add to the cognitive costs. The well known
phenomenon of ``ad blindness" and the wide-spread installation of ad blockers
are consumer counter-reactions to irrelevant ads.

Good ranking, which is necessarily personalized and based on comprehensive query
and document understanding, reduces the Delphic costs, including cognitive,
interactivity, and time costs. But our second thesis is that algorithmic ranking
only partially determines search utility. To wit, over the quarter century of
web search history, many reductions of Delphic costs were attained by mechanisms
unrelated to ranking, but affecting the human-machine interaction and the
quality of results. For example

\begin{itemize} \item Spelling correction, synonym substitution, query
completion, and query suggestions lower the costs of searching. Searchers
respond by self-correcting less often, further increasing the value of these
technologies.

\item Assiduous corpus curation (freshness, comprehensiveness, types of content,
safety checks) mitigates content impairments regardless of ranking.

\item Deeper understanding and more sophisticated processing of  content
allow more accurate query matching thus better ranking, but also better
snippets, cross-lingual search, and direct navigation to the relevant part of
the matching page or video thus reducing the cognitive and interactivity costs.

\item Fast and geographically distributed server farms reduce elapsed time and
importantly, encourage user query reformulation.

\item Diversity in results (rather than pure ranking maximization) increases the
probability of answering ambiguous queries, increasing the Delphic costs for
some and reducing for others. Additional user-feedback controls, e.g.~``More
like this" or ``More from this site" buttons, make it easy for the user to
reduce diversity when desired, thus reducing Delphic costs overall.

\item For more than a century it has been said that
\href{https://en.wikipedia.org/wiki/A_picture_is_worth_a_thousand_words}{\uline{``[a]
picture is worth 1000 words"}}~\cite{noauthor_picture_2023}. Early in the web history, search result pages
were almost entirely text, but in time they increasingly incorporated
multi-media, thus reducing both the interactivity and cognitive load while
raising the time and bandwidth costs. In the reverse direction ``a camera is the
new keyboard" in many contexts (e.g.~Pinterest lens, Bing visual search, Google
lens). Similarly, voice search reduces the time and effort to formulate a query,
especially when no keyboard is available.

\item Modern search engine result pages include types of content that do not
involve ranking at all, such as information extracted from databases or
knowledge graphs, information capsules purposely authored by experts, and answers to
related questions, all buttressed by the ability to select news, images, videos,
shopping, academic publications, and other presentation methods. A significant
percentage of Google and Bing results include such extracted information or
``rich features". All of these mechanisms reduce the Delphic costs of the users
who might be able to find what they need (or reformulate their queries) without
any further processing of the traditional results.

\item Users develop search strategies to reduce their personal Delphic costs.
Such strategies are persistent and Delphic cost reduction faces a trade-off
between optimizing for existing strategies and promoting new, more efficient
ones.
\end{itemize}

The trend towards facilitating task completion coupled  on one hand with major scientific  advances in Natural Language Processing and Voice Recognition, and on the other hand, the explosive proliferation of smart phones led in early 2010's to the emergence of
\href{https://en.wikipedia.org/wiki/Virtual_assistant}{\uline{virtual assistants}}~\cite{noauthor_virtual_2023},
such as Amazon's Alexa, Apple's Siri, Google's Assistant, Microsoft's Cortana, and Samsung's Bixby.   
These assistants use voice as a primary mode of interaction, and sometimes are known as ``voice assistants''. Although they offer some limited  search capabilities, in particular factoid elicitation, their main goal is  to complete a task, either on the mobile device (e.g. add a reminder, set an alarm, send a text, play some music) or via a connected ``smart'' devices like a thermostat, window shade, security alarm or door lock. 

From the point of view of Delphic costs, the voice interface that characterizes virtual assistants reduce  interactivity costs for very short queries but make long queries  difficult and in either case the risk of miscommunication is quite high.   The task completion capabilities are popular but for now the underlying tasks are rather simple so the total time savings are limited.  
Nevertheless, according to E-marketer~\cite{noauthor_how_nodate}, in 2022 about 42\%\ of the US population reported using a voice assistant in the last month.

Over the past year, there has been a dramatic growth in the use of
Generative and Interactive AI ("chatbots") for search, either as complements to
traditional web search engines (e.g.~Google, ``New Bing") or as standalone
solutions (e.g.~OpenAI's ChatGPT, Anthropic’s Claude). This is a rapidly changing
landscape; nevertheless the impact of chatbots on Delphic costs is obvious, extensive,
and offers a good explanation for the chatbots' immediate success.

Some features of chatbots increase Delphic benefits and reduce Delphic
costs:
\begin{itemize} 
\item Chatbots can save time and effort both on the  crafting of the query and on the processing of results. On input, chatbots enable users to express their intent in a natural form followed by interactive emendations, reducing both the cognitive and interactivity costs of classical web search.  On  output, for certain queries, chatbots provide short responses that embody knowledge synthesized from numerous and/or voluminous documents. This again significantly reduces cognitive and interactivity overhead, leading to substantial time savings, and overall higher  utility. 

\item Chatbots have unprecedented and growing capabilities to facilitate task completion.  For instance. while a classic search engine might offer advice on how to implement bubblesort in python, a chatbot can produce the code, ready for cut and paste.  Similarly chatbots are able to write stories, poems, novels, and screenplays, and create images, spreadsheets, and presentations.  Even though in most cases, further editing is required, the task facilitation is tremendous.

\item Chatbots have the potential to provide
\href{https://doi.org/10.1145/3476415.3476428}{\uline{expert answers}}~\cite{metzler_rethinking_2021} even to
naive questions.

\item Chatbots are naturally better suited to voice interaction, both input and
output. This is convenient, although it might increase the time to task
completion, since for many people typing and reading are faster than speaking
and listening.

\item The output can be specified to match the user needs and reduce the risk of
misunderstandings (e.g.~``explain in simple english", ``describe to an expert",
``in a few words", etc) and since no navigation is needed outside of the search
engine, the privacy and safety risks are arguably lower.

\item Chatbots offer potential hedonic benefits; in simpler words, at least for now, many
people are having fun interacting with chatbots whether seriously or
\href{https://dl.acm.org/doi/abs/10.1145/3491101.3519870}{\uline{facetiously}}~\cite{shani_alexa_2022}.

\end{itemize} Other features of chatbots increase Delphic costs or reduce the
benefits:

\begin{itemize} \item The access costs will likely increase, since the cost of
operating a chatbot is comparatively higher~\cite{mok_chatgpt_nodate}. It is unclear if chatbots can be
supported by advertising: currently Google and ``New Bing" are ``free'', but the
premium version of ChatGPT requires a subscription.

\item The latency of the answer is typically much higher than for web search.
This is somewhat masked by chatbots ``typing" their answer to simulate a human
interaction, but still apparent.

\item Currently the risk of misinformation is very high, due to
``hallucinations" and training data issues, in particular substantial delays
(relative to web search) in incorporating the latest information. For many
people this leads to distrust or even discomfort. Furthermore if every answer
needs to be checked (as some chatbots suggest) the time gains vanish.

The very polished look of the chatbots output might induce a false perception of reliability:  in a notorious case, a highly experienced lawyer accepted citations to a bunch of non-existent cases \cite{weiser_chatgpt_2023} since they seemed entirely plausible.   

\end{itemize} 

\begin{center}
    *
\end{center}

We hope that this paper will engender new ideas for Delphic costs
assessments, the measurement of Delphic costs, and means of reducing these
costs. We would like to see the evaluation of web search engines move away from
assessing the quality of ranking in isolation of the users’ overall search
experience and personal context towards a holistic evaluation of user utility
from using search engines. Moreover, this ``utilitarian analysis" approach,
rather than pure relevance analysis, could and should be applied to situations
that do not involve explicit search, such as content feeds and recommender
systems.

\section*{Acknowledgments}

\begin{enumerate} \item We benefited from discussions and feedback from numerous
colleagues; among them Alex Fabrikant, Alvin Roth, Bhargav Kanagal, Bruno
Possas, Chris Mah, Evgeniy Gabrilovich, Fernando Pereira, Ivan Kuz\-ne\-tsov,
Katrina Ligett, Marc Najork, Martin Abadi, Mike Bendersky, Mukund Sundararajan,
Prabhakar Raghavan, Robin Dua, Ronny Lempel, and Sandeep Tata. We thank them
all.

\item This paper represents the opinion of the authors, and does not necessarily
reflect the positions of Google LLC. \end{enumerate}

\defbibnote{excuse}{In this preliminary version the reference section is incomplete.}
\printbibliography[prenote=excuse]
\end{document}